\documentclass[twocolumn,showpacs,preprintnumbers,amsmath,amssymb]{revtex4}
\usepackage{graphicx}
\usepackage{dcolumn}
\usepackage{bm}
\usepackage[american]{babel}
\usepackage{enumerate}
\begin{document}

\title{First results from the new PVLAS apparatus: a new limit on vacuum magnetic birefringence}

\author{F.~Della Valle}
\affiliation{INFN, Sezione di Trieste and Dipartimento di Fisica, Universit\`a di Trieste, Via Valerio 2, I-34127 Trieste, Italy}
\author{A. ~Ejlli}
\affiliation{INFN, Sezione di Ferrara and Dipartimento di Fisica, Universit\`a di Ferrara, Polo Scientifico, Via Saragat 1 C, I-44100 Ferrara, Italy}
\author{U.~Gastaldi}
\affiliation{INFN, Sezione di Ferrara, Polo Scientifico, Via Saragat 1 C, I-44100 Ferrara, Italy}
\author{G.~Messineo}
\affiliation{INFN, Sezione di Ferrara and Dipartimento di Fisica, Universit\`a di Ferrara, Polo Scientifico, Via Saragat 1 C, I-44100 Ferrara, Italy}
\author{E.~Milotti}
\affiliation{INFN, Sezione di Trieste and Dipartimento di Fisica, Universit\`a di Trieste, Via Valerio 2, I-34127 Trieste, Italy}
\author{R.~Pengo}
\affiliation{INFN, Laboratori Nazionali di Legnaro, Viale dell'Universit\`a 2, I-35020 Legnaro}
\author{L.~Piemontese}
\affiliation{INFN, Sezione di Ferrara and Dipartimento di Fisica, Universit\`a di Ferrara, Polo Scientifico, Via Saragat 1 C, I-44100 Ferrara, Italy}
\author{G.~Ruoso}
\affiliation{INFN, Laboratori Nazionali di Legnaro, Viale dell'Universit\`a 2, I-35020 Legnaro}
\author{G.~Zavattini}
\affiliation{INFN, Sezione di Ferrara and Dipartimento di Fisica, Universit\`a di Ferrara, Polo Scientifico, Via Saragat 1 C, I-44100 Ferrara, Italy}
\collaboration{PVLAS Collaboration}
\noaffiliation

\date{\today}

\begin{abstract}
Several groups are carrying out experiments to observe and measure vacuum magnetic birefringence, predicted by Quantum Electrodynamics (QED). We have started running the new PVLAS apparatus installed in Ferrara, Italy, and have measured a noise floor value for the unitary field magnetic birefringence of vacuum $\Delta n_u^{\rm (vac)}= (4\pm 20) \times 10^{-23}$ T$^{-2}$ (the error represents a 1$\sigma$ deviation). This measurement is compatible with zero and hence represents a new limit on vacuum magnetic birefringence deriving from non linear electrodynamics. 
This result 
reduces to a factor 50 the gap to be overcome to measure for the first time the value of $\Delta n_u^{\rm (vac,QED)}$ predicted by QED: $\Delta n_u^{\rm (vac,QED)}= 4\times 10^{-24}$~T$^{-2}$. 
These birefringence measurements also yield improved model-independent bounds on the coupling constant of axion-like particles to two photons, for masses greater than 1 meV, 
 along with a factor two improvement of the fractional charge limit on millicharged particles (fermions and scalars), including neutrinos.
\end{abstract}


\maketitle

\section{Introduction}
Non linear electrodynamic effects in vacuum have been predicted since the earliest days of Quantum Electrodynamics (QED), a few years after the discovery of positrons \cite{QED,Dirac,Anderson}. One such effect is vacuum magnetic birefringence \cite{Adler}, closely connected to elastic light-by-light interaction. The effect is extremely small and has never yet been observed directly. 

Although today QED is a very well tested theory, the importance of detecting light-by-light interaction remains. Firstly QED has been tested always in the presence of charged particles in the initial state and/or the final state. No tests exist in systems with only photons. More in general, no interaction has ever been observed directly between gauge bosons present in both the initial and final states. Secondly, 
to date, the evidence for zero point quantum fluctuations relies entirely on the observation of the Casimir effect, which applies to photons only. Here we are dealing with the fluctuations of virtual charged particle-antiparticle pairs (of any nature, including hypothetical millicharged particles) and therefore the structure of fermionic quantum vacuum: to leading order, it would be a direct detection of loop diagrams. Finally, the observation of light-by-light interaction would be an evidence of the breakdown of the superposition principle and of Maxwell's classical equations.
One important consequence of a nonlinearity is that the velocity of light would now depend on the presence or not of other electromagnetic fields.

In a general framework of non-linear electrodynamics at the lowest order described by a Lorentz invariant parity conserving Lagrangian correction \cite{Denisov}
\begin{eqnarray}
L_{\rm nl}& =& \frac{\xi}{2\mu_{0}}\bigg[\eta_{1}\left(\frac{E^2}{c^2}-B^2\right)^2+4\eta_{2}\left(\frac{\vec{E}}{c}\cdot\vec{B}\right)^2\bigg],
\label{Lpm}
\end{eqnarray}
the induced birefringence due to an external magnetic field perpendicular to the propagation direction of light is
\begin{equation}
\Delta n^{\rm (vac)} = 2\xi\left(\eta_2-\eta_1\right) B^2 = \Delta n_u^{\rm (vac)} B^2.
\label{vacuumbirif}
\end{equation}
Here $\xi=1/B_{\rm crit}^{2}$ ($B_{\rm crit}={m^{2}c^{2}}/{e \hbar}=4.4\cdot10^{9}$~T), and $\eta_{1}$ and $\eta_{2}$ are dimensionless parameters depending on the chosen model.
In analogy to what is done for the Cotton-Mouton effect (for a review see Ref. \cite{Bishop}), we have defined the unitary field magnetic birefringence of vacuum $\Delta n_u^{\rm (vac)}$. Moreover vacuum magnetic birefringence due to axion-like particles (ALP) and millicharged particles also depends on $B^2$ \cite{Cameron,PRD,Ni1996,Ni2010,pugnat}. These last two hypothetical effects represent new physics beyond the Standard Model and can be searched for in a model independent way with an apparatus such as PVLAS. In particular, ALPs in the mass range up to 1 meV have long been considered as cold dark matter candidates \cite{wisp}.

In the Euler-Heisenberg electrodynamics $\eta_{1}^{\rm (QED)}=\frac{4}{7}\eta_{2}^{\rm (QED)}=\alpha/(45\pi)$, $\alpha={e^2}/{(\hbar c 4\pi\epsilon_0)}$ being the fine structure constant. In this case
\begin{equation}
 \Delta n_u^{\rm (vac,QED)}= \frac{2}{15\mu_{0}}\frac{\alpha^2 \mathchar'26\mkern-10mu\lambda_e^{3}}{m_{e}c^{2}}=3.97\times10^{-24} {\text{~T}}^{-2}
\label{birifqed}
\end{equation}


The ellipticity $\psi$ induced on a beam of linearly polarized laser light of wavelength $\lambda$ which traverses a vacuum region of length $L$, where a magnetic field $B$ orthogonal  to the direction of light propagation is present, is given by \cite{Iacopini,HypIn,PRD78,NPJ,NIMA}
\begin{equation}
\psi =  \frac{N \pi \Delta n_u^{\rm(vac)}\int_0^L{ B^2 dl }}{\lambda} \sin2\vartheta
\label{ellitticita}
\end{equation}
where $\vartheta$ is the angle between the directions of the polarization vector and of the magnetic field vector and  $N$ is the number of times the medium is traversed by the light.

An ellipsometric method to observe vacuum magnetic birefringence was proposed by E. Iacopini and E. Zavattini in 1979 \cite{Iacopini}. Experimental attempts started in the nineties \cite{Cameron,HypIn} and several are ongoing \cite{PRD78,NPJ,NIMA,bmv,Ni2010,pugnat}. The Lagrangian (\ref{Lpm}) also predicts direct light-light elastic scattering. 
See Refs.\cite{Bernard,BernardOld,Brodin,exawatt,Luiten1,Milotti} for experimental attempts. Neither method has reached the capability of detecting this fundamental nonlinear effect regarding light by light interaction. Presently published results on $\Delta n_u^{\rm(vac)}$ determined from ellipsometric experiments are reported in Table \ref{limits}.
\begin{table}[h!]
\caption{Presently published results on $\Delta n_u^{\rm(vac)}$ obtained from ellipsometric measurements. Values are in $10^{-23}~{\rm T}^{-2}$.}
\begin{tabular}{l|c|c|c}
\hline\noalign{\smallskip}
Experiment & Central value  &$1\;\sigma$  & Refs.\\
\noalign{\smallskip}\hline\noalign{\smallskip}
 BFRT & 22000 & 2400 & \cite{Cameron}\\
PVLAS - LNL & 640 & 780 & \cite{PRD78}\\
PVLAS - FE test setup & 840 & 400 & \cite{NPJ}\\
BMV & 830 & 270 & \cite{bmvnew}\\
\noalign{\smallskip}\hline
\end{tabular}
\label{limits}
\end{table}

In this letter we report on a significant improvement obtained after the commissioning of the new PVLAS experimental setup installed at the INFN section of Ferrara. 
The principle of the experiment is explained in \cite{Iacopini,NPJ}. 
The calibration of the apparatus has been done by measuring the Cotton-Mouton effect of $\rm O_2$ and $\rm He$ gases at low pressures and controlling their consistency with the values present in literature. 
In this paper we briefly summarize the main features of the new experimental setup and focus on the measurements giving a new conservative upper limit on $\Delta n_u^{\rm (vac)}$. 

\section{Experimental Method and Apparatus}
\begin{figure}[htb]
\centering\includegraphics[width=\linewidth]{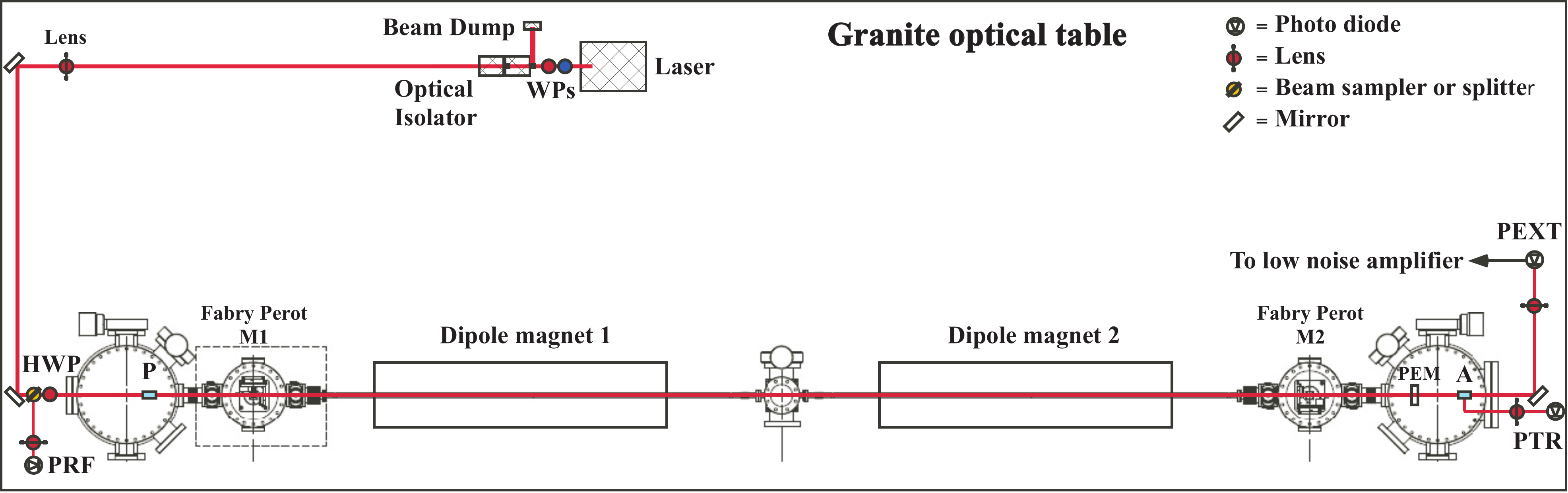}\\
\centering\includegraphics[width=\linewidth]{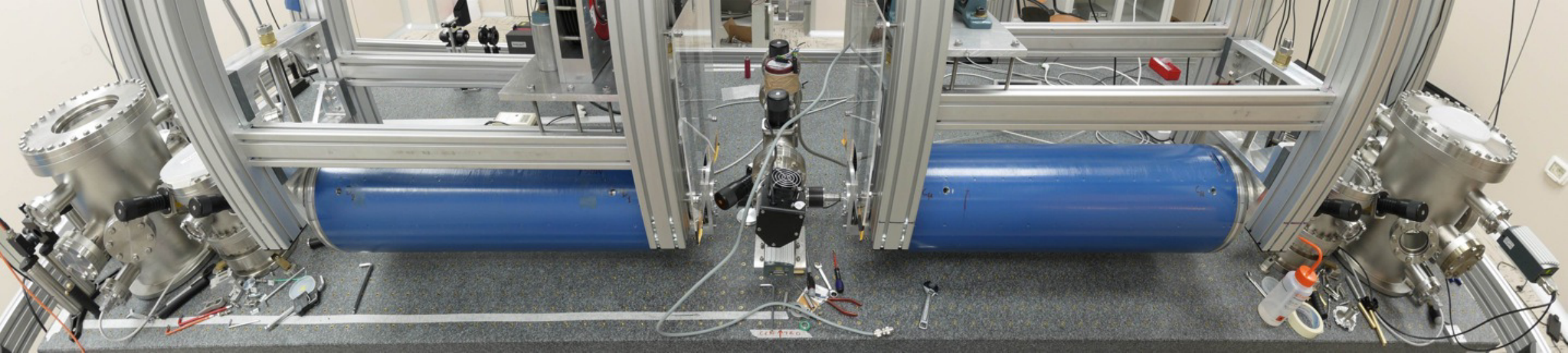}
\caption{Upper panel: Scheme of the apparatus. The granite optical table,  4.8 m x 1.5 m, is shown together with the optical components and the five vacuum chambers. 
{\bf HWP} = Half wave plate; {\bf P} = Polarizer; {\bf A} = Analyzer;  {\bf WPs} = Wave plates; {\bf PEM} = Photoelastic Modulator; {\bf PRF} = Reflection photodiode; {\bf PTR} = Transmission photodiode; {\bf PEXT} = Extinction photodiode. Lower panel: A wide-angle picture of the PVLAS apparatus. The two blue cylinders are the permanent magnets: they  are hanging from an aluminium structure mechanically decoupled from the rest of the optical table.}
\label{apparato}
\end{figure}

The upper and lower panels of figure \ref{apparato} show a schematic top view and a photograph of the apparatus. Tables \ref{optics}, \ref{vacuum}, and \ref{magnets} summarize the parameters and performances of its different components.
\begin{table}[h!]
\caption{Optics characteristics.}
\begin{tabular}{l|l}
\hline\noalign{\smallskip}
Optical bench & Material: granite\\
& Weight: 4.5 tons\\
& Dimensions: $4.8 \times 1.5 \times 0.5$~m$^3$\\
\noalign{\smallskip}\hline\noalign{\smallskip}
 Laser & Solid state Nd:YAG NPRO\\
  & Wavelength: $\lambda = 1064$~nm\\
 & Maximum power: 2~W\\
\noalign{\smallskip}\hline\noalign{\smallskip}
Fabry-Perot & Length: 3.303~m\\
& Finesse: 670000\\
\noalign{\smallskip}\hline\noalign{\smallskip}
Fabry-Perot &  Radius of curvature: 2~m\\
mirrors &  Reflectivity: $R > 0.999996$ \\
&  Transmission: $T\approx 2\times 10^{-6}$\\
\noalign{\smallskip}\hline\noalign{\smallskip}
Photo Elastic& Fused silica bar coupled \\
  Modulator ${\bf PEM}$& to a piezo transducer.\\
  &Resonance frequency: 50.047 kHz\\
& Typically induced ellipticity: $\approx 10^{-3}$\\
\noalign{\smallskip}\hline\noalign{\smallskip}
Polarizers  & Nominal extinction ratio $\sigma^2 < 10^{-7}$\\
${\bf P}$ and ${\bf A}$& 1 cm clear aperture\\
\noalign{\smallskip}\hline
\end{tabular}
\label{optics}
\end{table}
\begin{table}[h!]
\caption{Vacuum system characteristics.}
\begin{tabular}{l|l}
\hline\noalign{\smallskip}
Components& Non magnetic materials \\
& Turbomolecular pumps\\
&  Non-evaporable getter pumps\\
\noalign{\smallskip}\hline\noalign{\smallskip}
Total pressure & $\lesssim 10^{-7}$~mbar, mainly H$_2$O.\\
\noalign{\smallskip}\hline
\end{tabular}
\label{vacuum}
\end{table}
\begin{table}[h!]
\caption{Magnet characteristics.}
\begin{tabular}{l|l}
\hline\noalign{\smallskip}
Components & 2 permanent dipole magnets in\\
 &  Halbach configuration.\\
& Central bore 20 mm\\
& Physical length 96 cm \\
& External diameter 28 cm\\
& Weight 450 kg\\
& Include a magnetic field shielding.\\
\noalign{\smallskip}\hline\noalign{\smallskip}
Field strength & $B_{\rm max} = 2.6$~T\\
& $\int{B^2 dl} = 5.12$~T$^{2}$m each.\\
& Stray field $< 1$ gauss\\
& (along axis @ $20 $~cm).\\
\noalign{\smallskip}\hline\noalign{\smallskip}
Rotation frequency & Up to 10 Hz.\\
\noalign{\smallskip}\hline
\end{tabular}
\label{magnets}
\end{table}

Linearly polarized laser light  is injected into the ellipsometer which is installed inside a high vacuum enclosure. The ellipsometer consists of an entrance polarizer {\bf P} and an analyser {\bf A} set to maximum extinction. Between {\bf P} and {\bf A} are installed the entrance mirror {\bf M1} and the exit mirror {\bf M2} of a Fabry-Perot cavity {\bf FP} with ultra-high finesse ${\cal F}$ \cite{recordfinesse}. The light back-reflected by the {\bf FP} is detected by the photodiode {\bf PRF}, and is used by a feedback system which locks the laser frequency to the {\bf FP} with a variant of the Pound-Drever-Hall technique \cite{rsi95}. The resonant light between the two mirrors traverses the bore of two identical permanent dipole magnets (see Table \ref{magnets}). The magnets can rotate around the {\bf FP} cavity axis so that the magnetic field vectors of the two magnets rotate in planes normal to the path of the light stored in the cavity.  The motors driving the two magnets are controlled by two phase locked signal generators. The same signal generators trigger the data acquisition. The magnetic field of the magnets induce a birefringence on the medium in the bores; the {\bf FP} enhances the ellipticity acquired by the light by a factor $N = 2{\cal F}/\pi$.  Due to the rotation of the magnetic field, the induced ellipticity varies harmonically at twice the rotation frequency of the magnets [see the dependence of $\psi$ from $2\vartheta$ in equation (\ref{ellitticita})]. Given the parameters of our apparatus ($\lambda = 1064$~nm, $\int{B^2 dl} = 10.25$~ T$^2$m) the predicted ellipticity generated by vacuum magnetic birefringence after a single passage of the light through the magnets is $\psi_{\rm single} = 1.2 \times  10^{-16}$. The {\bf FP} cavity multiplies the single pass ellipticity $\psi_{\rm single}$ by a factor $N = 4.3\times 10^5$ resulting in an ellipticity to be measured of $\psi^{\rm (vac)} = 5 \times 10^{-11}$.

A photo elastic modulator ({\bf PEM}) then adds a known small ellipticity at a fixed frequency $\Omega_{\rm PEM}$. Under these conditions the intensity $I_{\rm out}(t)$ of the light emerging from the analyzer {\bf A} is    
\begin{eqnarray}
& I_{\rm out}(t)= I_{0}\left[\sigma^2+\left|\imath\eta(t)+\imath\psi\sin{2\vartheta(t)}+\imath\alpha(t)\right|^{2}\right]\simeq\nonumber\\
&\simeq I_{0}\left[\sigma^2 +\eta(t)^2+2\eta(t)\psi\sin2\vartheta(t)+2\eta(t)\alpha(t)\right]
\end{eqnarray}
where $I_{0}$ represents the light power reaching the analyser, $\eta(t)$ the ellipticity modulation generated by the {\bf PEM}, $\sigma^2$ is the extinction ratio of the two polarizers and $\alpha(t)$ describes the slowly varying spurious ellipticities present in the apparatus.
As can be seen, the introduction of the {\bf PEM} linearises the ellipticity signal which would otherwise be quadratic. The light emerging from the analyser is collected by the photodiode {\bf PEXT}.  

The most important Fourier components of $I_{\rm out}(t)$ come from the terms $2\eta(t)\psi\sin2\vartheta(t)$ and $\eta(t)^2$. The first of these terms results in the beating of the ellipticity induced by the {\bf PEM} (at $\Omega_{\rm PEM}$) and the ellipticity induced by the rotating magnets (at $2\Omega_{\rm Mag}$). 
 The term  $\eta(t)^2$ generates a Fourier component at $2\Omega_{\rm PEM}$. 

During acquisition the photodiode signal coming from {\bf PEXT} is therefore demodulated at the frequency $\Omega_{\rm PEM}$ 
and at its second harmonic $2\Omega_{\rm PEM}$. Both these demodulated signals, respectively $I_{\Omega_{\rm PEM}}(t)$ and $I_{2\Omega_{\rm PEM}}(t)$, are acquired by a data acquisition system together with the ordinary beam intensity $I_{0}$ exiting the analyser {\bf A}. With the DC component of $I_{2\Omega_{\rm PEM}}(t)$, indicated as  $I_{2\Omega_{\rm PEM}}(\rm DC)$, and $I_{\Omega_{\rm PEM}}(t)$
the ellipticity signal $\psi(t)$ can be determined by the equation
\begin{equation}
\psi(t) = \frac{I_{\rm \Omega_{\rm PEM}}(t)}{\sqrt{8 I_0 I_{\rm 2\Omega_{\rm PEM}}(\rm DC)}}.
\label{ellipticity}
\end{equation}
With the magnets rotating at $\Omega_{\rm Mag}$, a magnetically induced birefringence will generate a Fourier component of $\psi(t)$ at $2\Omega_{\rm Mag}$. 

Magnetic field sensors and laser locking signals are also acquired to determine the phase of $\psi(t)$ and for diagnostics. These signals are sampled at 32 times the rotation frequency of the magnets (typically 3 Hz) by a 16 bit multi channel ADC board.

The vacuum system must guarantee that the presence of residual gas species do not mask vacuum magnetic birefringence. Indeed the Cotton-Mouton effect induces a magnetic birefringence in gases which depends on $B^2$ exactly like vacuum magnetic birefringence. The magnetic birefringence of gases also depends linearly on pressure. In Table \ref{cm} the equivalent partial pressures  $P_{\rm eq}$ which would mimic a vacuum magnetic birefringence for various gases \cite{Bishop,cmHe,cmHeRizzo,cmh2o} are reported. The vacuum system must maintain these species well below their vacuum equivalent pressures.
\begin{table}[h!]
\caption{Vacuum equivalent pressures  $P_{\rm eq}$ for various gases.}
\begin{tabular}{c|c|c}
\hline\noalign{\smallskip}
Gas & $\Delta n_u$ [T$^{-2}$atm$^{-1}$] & $P_{\rm eq}$ [mbar]\\
\noalign{\smallskip}\hline\noalign{\smallskip}
He & $2.1\times 10^{-16}$ & $2 \times 10^{-5}$\\
Ar & $7\times 10^{-15}$ & $6 \times 10^{-7}$\\
H$_2$O & $6.7 \times 10^{-15}$ &$6 \times 10^{-7}$ \\
CH$_4$ & $1.6 \times 10^{-14}$ &$3 \times 10^{-7}$\\
O$_2$ & $-2.5 \times 10^{-12}$ & $2 \times 10^{-9}$\\
N$_2$ & $-2.5 \times 10^{-13}$ & $2 \times 10^{-8}$\\
\noalign{\smallskip}\hline
\end{tabular}
\label{cm}
\end{table}

\section{Calibration}
\begin{figure}[htb]
\centering\includegraphics[width=\linewidth]{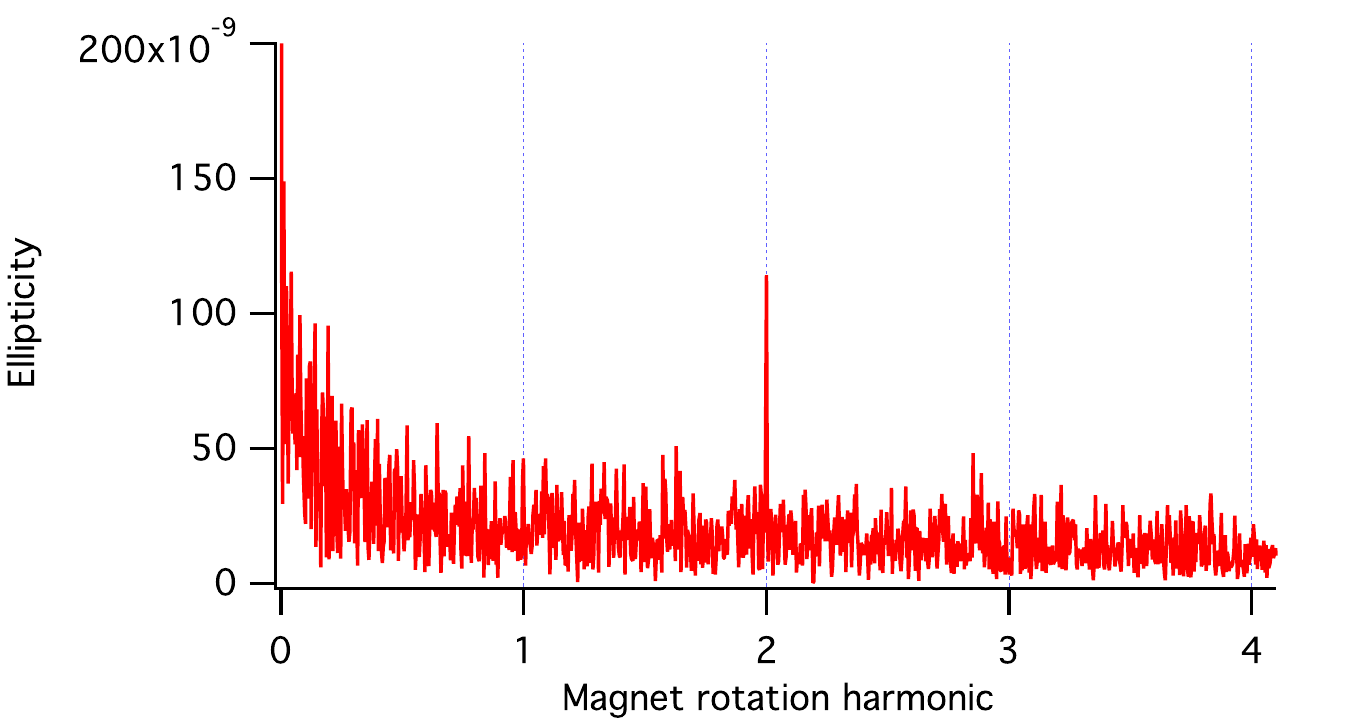}
\caption{Fourier spectrum of the measured ellipticity $\psi(t)$ with 32 $\mu$bar pressure of He.
The integration time was T = 4 hours. The peak at $2\Omega_{\rm Mag}$ corresponds to $\psi = 1.13 \times10^{-7}$. The vacuum magnetic birefringence predicted by QED is equivalent to a He pressure of $\sim 20$~nbar.}
\label{He32}
\end{figure}

\begin{figure}[h!]
\centering\includegraphics[width=\linewidth]{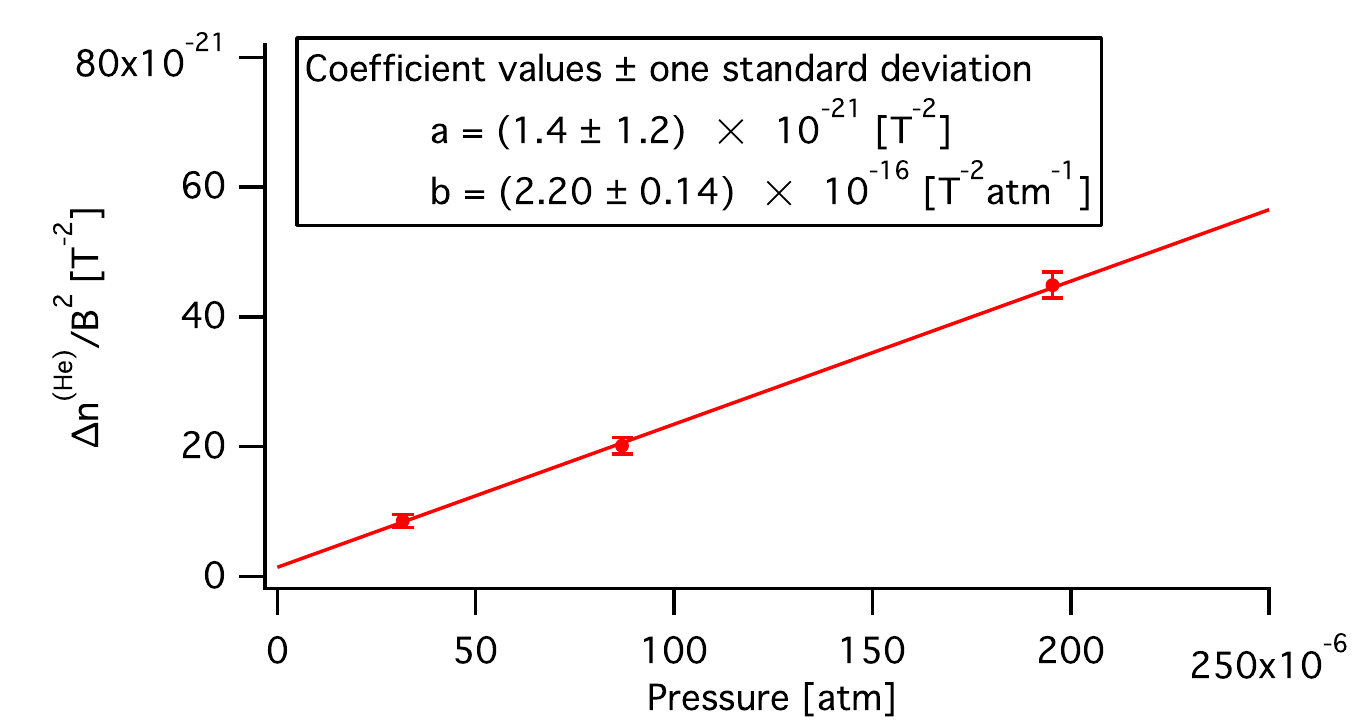}
\caption{Measured ${\Delta n^{\rm (He)}}/{B^2}$ as a function of pressure ${P}$. The error bars correspond to a 1$\sigma$ statistical error. The data are fitted with a linear function $a+b{P}$.}
\label{helium}
\end{figure}

Calibration of the apparatus is done using the Cotton-Mouton effect. In this case we used low pressure oxygen, which gives large signals. More importantly, we have also checked the calibration of the apparatus with low pressure helium, so as to induce a small ellipticity and demonstrate the sensitivity of the entire system. The lowest pressure of helium used was $P^{(\rm He)} = 32~\mu$bar. Considering that the unitary birefringence (B = 1~T and pressure = 1~atm) of helium due to the Cotton-Mouton effect is $\Delta n_{u}^{\rm (He)} = (2.1\pm0.1) \times 10^{-16}$~T$^{-2}$atm$^{-1}$ \cite{cmHe,cmHeRizzo}, the birefringence induced @ $ B = 2.5$~T and $ P = 32~\mu$bar is $\Delta n^{\rm (He)} = 3.9\times 10^{-20}$.
In figure \ref{He32} the Fourier transform of the measured ellipticity signal $\psi(t)$ is shown. There is a clear peak at $2\Omega_{\rm Mag}$, corresponding to an ellipticity of $(1.13\pm0.13)\times 10^{-7}$, with no spurious peaks present at other harmonics. The integration time was $T = 4$ hours. Given that ${\cal F} = 6.7\times 10^5$, $\lambda = 1064$~nm, $\int{B^2 dl} = 10.25$~T$^2$m, from the amplitude of the He peak at $32~\mu$bar,  the value of $\Delta n_u$ for helium results $\Delta n_u^{\rm (He,PVLAS)} = (2.2\pm0.1)\times 10^{-16}$~T$^{-2}$atm$^{-1}$, in perfect agreement with other published values \cite{Bishop,cmHe,cmHeRizzo}. It must be noted that this value is obtained from a single low pressure point. Other two low pressure points were also taken. Figure \ref{helium} shows a graph of $\Delta n^{\rm (He)}/B^2$ as a function of pressure ${P}$. 

The calibration process also allows the determination of the physical phase of the Fourier components: the ellipticity induced by a magnetic birefringence is maximum when the magnetic field is at $\vartheta = \pm45^{\circ}$ with respect to the polarization direction. Since a magnetic birefringence can be either positive or negative, the physical phase is determined $\mod 180^{\circ}$. A magnetically induced birefringence must have a phase consistent with the calibration phase. The ellipticity amplitudes determined from the Fourier transforms of the data obtained in vacuum are therefore projected along the physical axis and along the non-physical orthogonal axis.

\section{Results}
The data presented in this paper have been collected by rotating the two magnets at frequencies ranging from 2.4 Hz to 3 Hz for a total of 210 hours. Of these, 40 hours have been acquired with the magnets rotating at slightly different frequencies so as to check that neither of the two was generating spurious signals.

\begin{figure}[htb]
\centering\includegraphics[width=\linewidth]{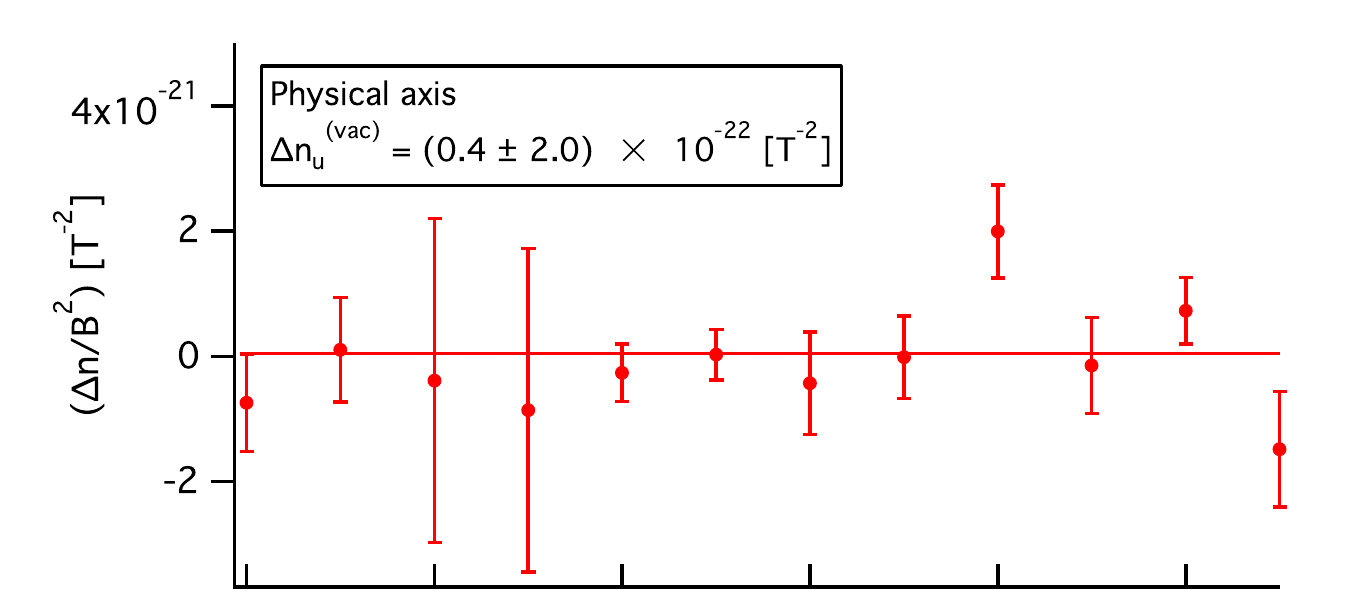}\\
\centering\includegraphics[width=\linewidth]{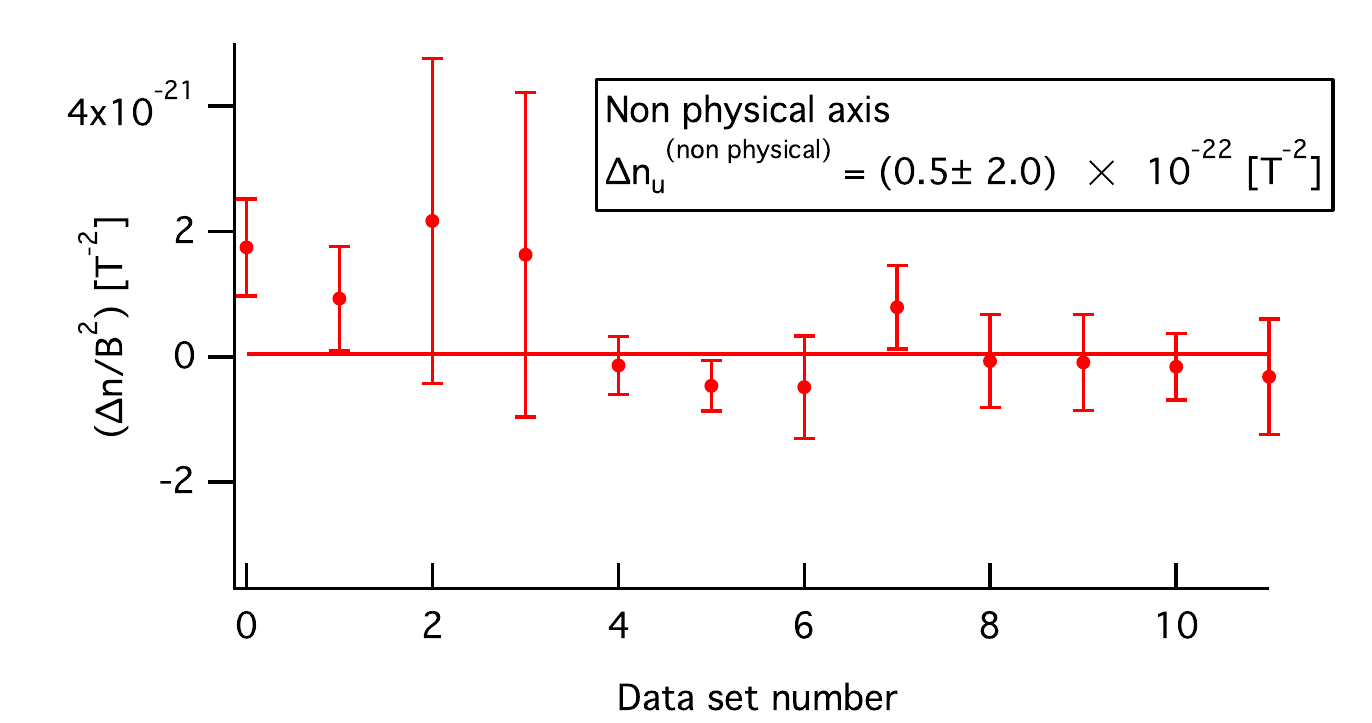}
\caption{Projections of $\Delta n/B^2$ along the physical and the non-physical axes for all the data sets. The horizontal red line represents the weighed average for all the runs.}
\label{risultati}
\end{figure}

The data analysis procedure is as follows:
\begin{enumerate}[1)]
\item For each run, lasting typically one day, the acquired signals are subdivided in blocks of 8192 points (256 magnet revolutions) and a Fourier transform of the ellipticity signal $\psi(t)$, calculated using equation (\ref{ellipticity}), is taken for each block.
\item For each block, the average noise in the ellipticity spectrum around $2\Omega_{\rm Mag}$ is taken. The ellipticity amplitude noise follows the Rayleigh distribution $P(\rho) = (\rho/\sigma^2) e^{-\frac{\rho^2}{2\sigma^2}}$, in which the parameter $\sigma$ represents the standard deviation of two identical independent Gaussian distributions for two variables $x$ and $y$ and $\rho = \sqrt{x^2+y^2}$. In our case $x$ and $y$ represent the projections of the ellipticity value at $2\Omega_{\rm Mag}$ along the physical and the non-physical axes. The average of $P(\rho)$ is related to $\sigma$ by $\langle P \rangle = \sigma\sqrt{\pi/2}$. For each data block, $\sigma$ is determined. This value is used in the next step as the weight for the ellipticity value at $2\Omega_{\rm Mag}$. 
\item For each run, a weighted vector average of the Fourier components of the ellipticities at $2\Omega_{\rm Mag}$, determined in step 2), is taken. 
\item Using the values for $\cal F$, $\int{B^2 dl}$ and $\lambda$ for each run, $\Delta n/B^2$ and $\sigma/B^2$ are determined. $\Delta n/B^2$ is then projected onto the physical and non-physical axes.
These values are plotted in figure \ref{risultati}.
\end{enumerate}
The weighted vector average of all the runs results in a value for the unitary birefringence of vacuum, with a 1$\sigma$ error, of
\begin{equation}
\Delta n_u^{\rm (vac)} \pm \sigma_{\Delta n_u^{\rm (vac)}} = (4\pm20)\times 10^{-23} {\rm~T}^{-2}
\label{deltanu}
\end{equation}
for the physical component (same phase and sign as for the helium Cotton-Mouton birefringence). For the non-physical component one finds 
$\Delta n_u^{\rm (non\;physical)} \pm  \sigma_{\Delta n_u^{\rm (vac)}}= (5\pm20)\times 10^{-23} {\rm~T}^{-2}$. 
This new limit is about a factor 50 from the predicted QED value of equation (\ref{birifqed}), $\Delta n_u^{\rm (vac,QED)}= 3.97 \times 10^{-24}{\rm~T}^{-2}$.

\section{Discussion and conclusions}
\subsection{QED}
\begin{figure}[htb]
\centering\includegraphics[width=\linewidth]{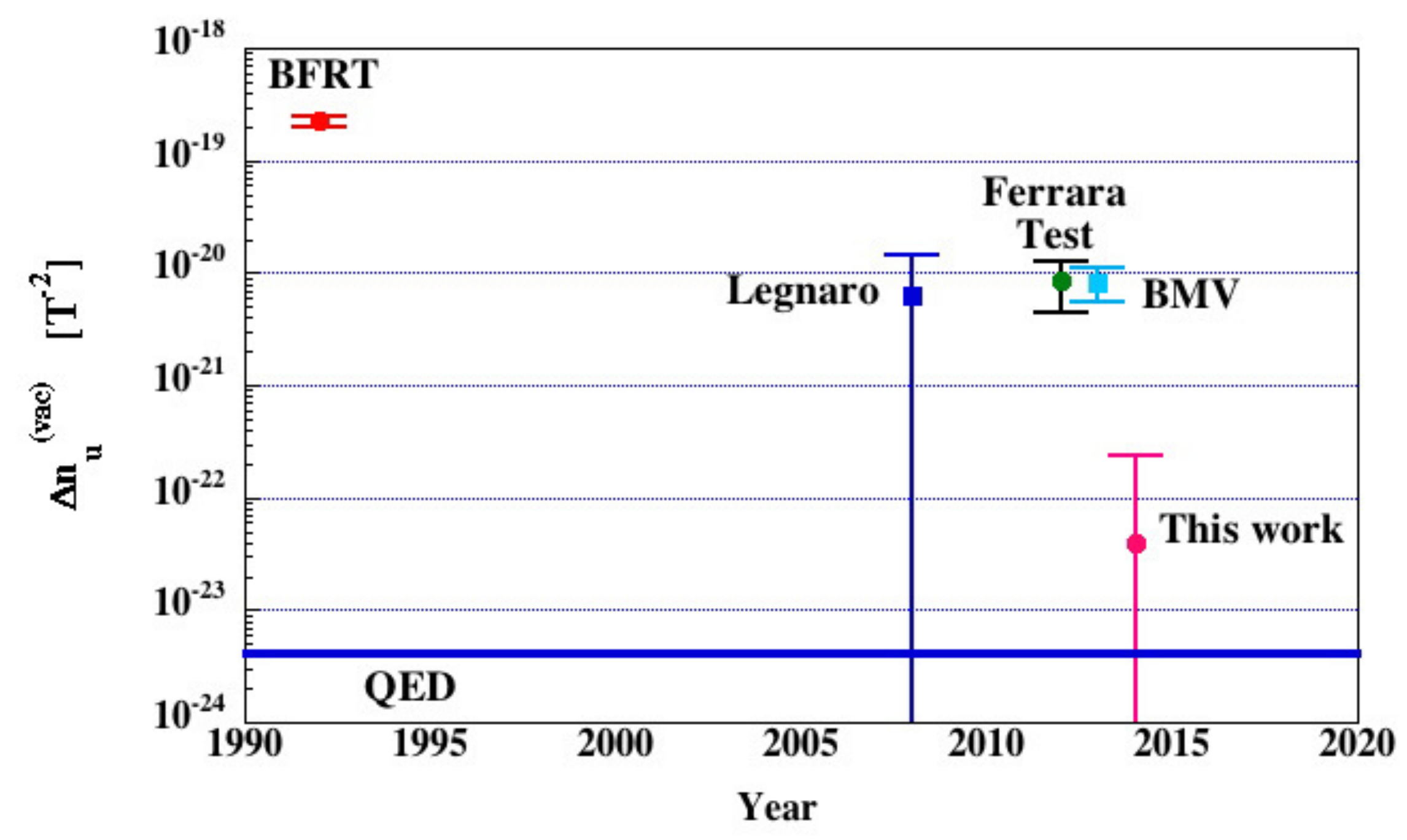}
\caption{Comparison of published results for $\Delta n_u^{\rm (vac)}$ of ellipsometric experiments (BFRT = \cite{Cameron}, Legnaro = \cite{PRD78}, Ferrara Test = \cite{NPJ}, BMV = \cite{bmvnew}). The error bars correspond to a $1\sigma$~c.l.}
\label{Flimiti}
\end{figure}

We have reported here on a significant improvement in the measurement of the magnetic birefringence of vacuum. In the Euler-Heisenberg framework we are now only a factor 50 away from the theoretical parameter, $\Delta n_u^{\rm (vac,QED)}=3.97\times10^{-24}$~T$^{-2}$, describing this effect. 
Our new limit is
\begin{equation}
\Delta n_u^{\rm (vac)} \pm \sigma_{\Delta n_u^{\rm (vac)}} = (4\pm20)\times 10^{-23} {\rm~T}^{-2}.
\end{equation}
In figure \ref{Flimiti} we compare previously published results with our new value and with the predicted QED effect. 

In the Euler-Heisenberg framework where $\eta_{1}^{\rm (QED)}=\frac{4}{7}\eta_{2}^{\rm (QED)}=\alpha/(45\pi)$, the elastic photon-photon total cross section for non polarized light depends directly on $\Delta n_u^{\rm (vac,QED)}$.
In the limit of low energy photons, $E_{\gamma} \ll m_{e}c^{2}$ \cite{DeTollis,Karplus,Duane},
\begin{equation}
\sigma_{\gamma\gamma}^{\rm (QED)}(E_{\gamma})=\frac{973\mu_{0}^{2}}{180\pi}\frac{E_{\gamma}^{6}}{\hbar^{4}c^{4}}(\Delta n_u^{\rm (vac,QED)})^{2}.
\end{equation}
From the experimental bound on $\Delta n_u^{\rm (vac)}$ one can therefore place an upper bound on $\sigma_{\gamma\gamma}^{\rm (QED)}$:
\begin{equation}
\sigma_{\gamma\gamma}^{\rm (QED)} < 4.6\times10^{-66} {\text {~m}}^2 {~@~1064~\rm nm}
\end{equation}
The QED prediction for this number is instead $\sigma_{\gamma\gamma}^{\rm (QED)} = 1.84\times 10^{-69}$~m$^2$. 

Although the sensitivity of our apparatus is far from its theoretical shot noise limit, integration in the absence of spurious peaks at the frequency of interest has allowed this significant improvement. The origin of the excess noise is still unknown but is clearly due to the presence of the Fabry-Perot cavity: without the cavity shot noise is achieved. We suspect that the origin of this noise is due to variations in the intrinsic birefringence of the reflective coating due to thermal effects.
Nonetheless at present the ellipsometric technique is the most sensitive one for approaching low-energy non-linear electrodynamics effects. 
Efforts will now go into the improvement of the sensitivity.
\subsection{Axion like particles}
Compared to model dependent constraints deriving from astrophysics \cite{Ringwald}, limits from laboratory experiments cannot compete. Nevertheless they can set new model independent limits on the coupling constant of axion-like particles (ALP) to two photons. In the results presented here only the runs with both magnets rotating at the same frequency were used, so that the total field length could be taken as the sum of the two magnet lengths. 

The magnetic birefringence induced by low mass axion-like particles can be expressed as \cite{Cameron}
\begin{equation}
\Delta n^{\rm (vac: ALP)}= \frac{g_{a}^{2}B^{2}}{2m_{a,s}^{2}}\left(1-\frac{\sin2x}{2x}\right)
\label{pseudo}
\end{equation}
where $g_{a}$ is the ALP - 2 photon coupling constant, $m_a$ its mass, $x=\frac{Lm_{a}^{2}}{4\omega}$, $\omega$ is the photon energy and $L$ is the magnetic field length. The above expression is in natural Heavyside-Lorentz units whereby 1~T $=\sqrt{\frac{\hbar^{3}c^{3}}{e^{4}\mu_{0}}}= 195$~eV$^2$ and 1~m $=\frac{e}{\hbar c}=5.06\times10^{6}$~eV$^{-1}$. 

In the approximation for which $x\ll1$ (small masses) this expression becomes
\begin{equation}
\Delta n^{\rm (vac: ALP)}= \frac{g_{a}^{2}B^{2}m_{a}^{2}L^{2}}{48\omega^2}
\end{equation}
whereas for $x \gg 1$
\begin{equation}
\Delta n^{\rm (vac: ALP)}= \frac{g_{a}^{2}B^{2}}{2m_{a}^{2}}.
\end{equation}
From our limit on $\Delta n_u^{\rm(vac)}$ given in equation (\ref{deltanu}) one can plot a new model independent exclusion plot for ALPs. Above $10^{-3}$~eV there is an improvement on the upper limit of $g_a$ with respect to previously published model independent limits.
\begin{figure}[h]
\centering\includegraphics[width=\linewidth]{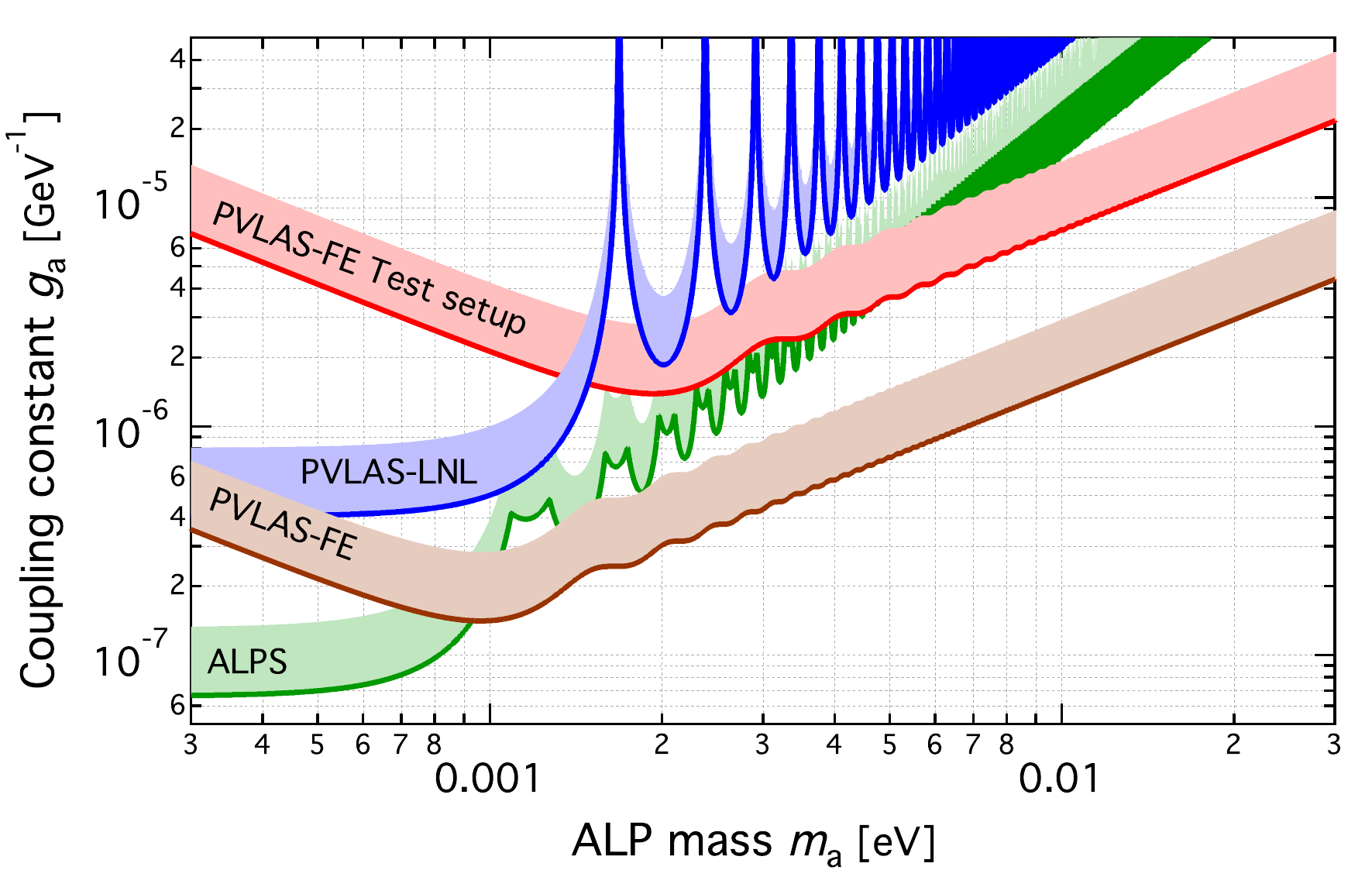}
\caption{Updated 95 \% c.l. exclusion plot for axion-like particles. In green, limits from the ALPS collaboration \cite{ALPs}; in blue, limits from dichroism measurements performed by PVLAS at LNL \cite{PRD}; in red, limits from the ellipticity measurements performed with the test setup in Ferrara \cite{NPJ}. The results described in this paper lead to a new bound, shown in brown. Preliminary results from the OSQAR collaboration can be found in Ref. \cite{osqar}. They are very similar to the results from ALPS.}
\label{alp}
\end{figure}
\subsection{Millicharged particles}

Slightly better exclusion plots can also be derived from $\Delta n_u^{\rm (vac)}$ for fermion and scalar millicharged particles. The vacuum magnetic birefringence due to the existence of such hypothetical millicharged particles can be calculated following \cite{Tsai1975,Daugherty1983,Ahlers2007}. By defining the ratio of the charge $q$ of such particles to the charge of the electron $\epsilon = q/e$ and $\chi$ as
\begin{equation}
\chi\equiv
\frac{3}{2}\frac{\hbar\omega}{m_{\epsilon}c^{2}}\frac{\epsilon e B\hbar}{m_{\epsilon}^{2}c^{2}}.
\label{chi}
\end{equation}
it can be shown that 
\begin{eqnarray}
\label{deltandf}
&&\Delta n^{\rm (vac: fermion)}=\\
&=&\left\{\begin{array}{ll}
\Delta n_u^{\rm (MCP)} B^{2}& (\chi \ll 1) \\
\displaystyle-\frac{135}{14}\frac{\pi^{1/2}2^{1/3}\left(\Gamma\left(\frac{2}{3}\right)\right)^{2}}{\Gamma\left(\frac{1}{6}\right)}\chi^{-4/3}\Delta n_u^{\rm (MCP)} B^{2}& (\chi \gg 1).
\end{array}\right.\nonumber
\end{eqnarray}
\begin{eqnarray}
\label{deltansc}
&&\Delta n^{\rm (vac: scalar)}=\\
&=&\left\{\begin{array}{ll}
\displaystyle-\frac{1}{2} \Delta n_u^{\rm (MCP)}  B^{2}& (\chi \ll 1) \\
\displaystyle\frac{135}{28}\frac{\pi^{1/2}2^{1/3}\left(\Gamma\left(\frac{2}{3}\right)\right)^{2}}{\Gamma\left(\frac{1}{6}\right)}\chi^{-4/3}\Delta n_u^{\rm (MCP)}  B^{2}&(\chi \gg 1).
\end{array}\right.\nonumber
\end{eqnarray}
\\
where, in analogy to QED, $\Delta n_u^{\rm (MCP)}$ is
\begin{equation}
\Delta n_u^{\rm (MCP)}=\frac{2}{15\mu_{0}}\frac{\epsilon^{4}\alpha^{2} \mathchar'26\mkern-10mu\lambda_\epsilon^{3}}{m_{\epsilon}c^{2}}
\end{equation}
\begin{figure}[h]
\centering\includegraphics[width=\linewidth]{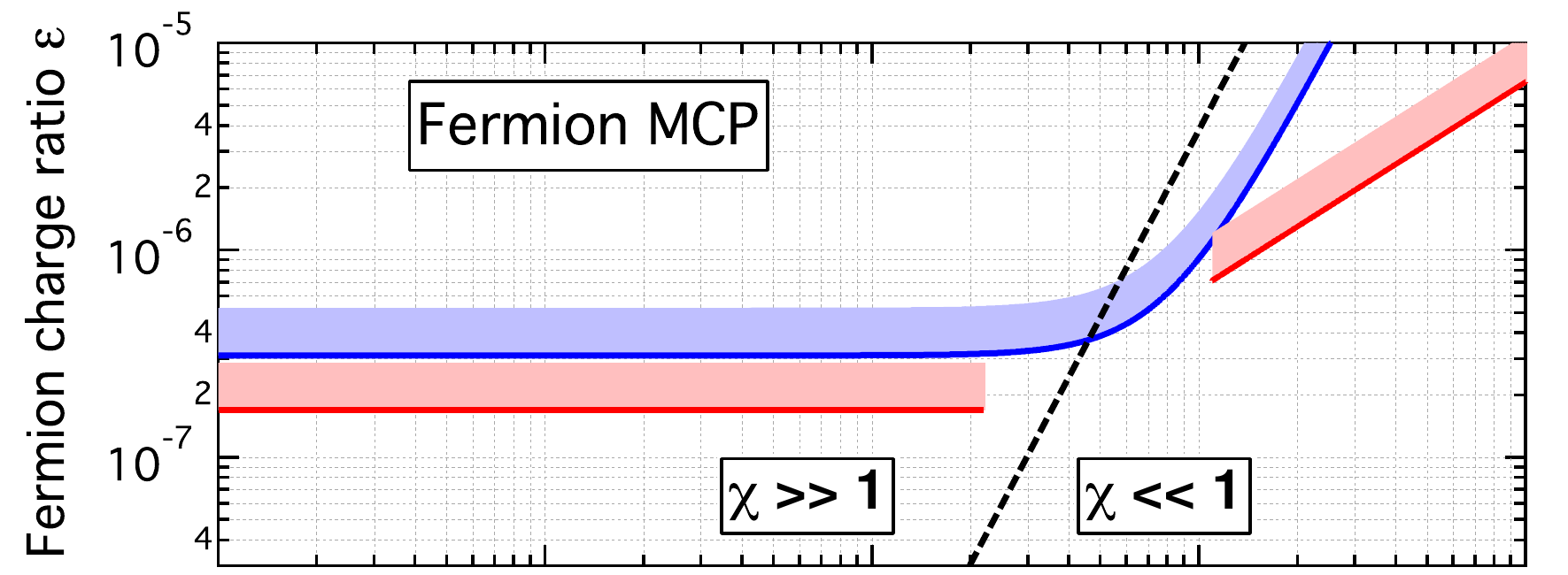}\\
\centering\includegraphics[width=\linewidth]{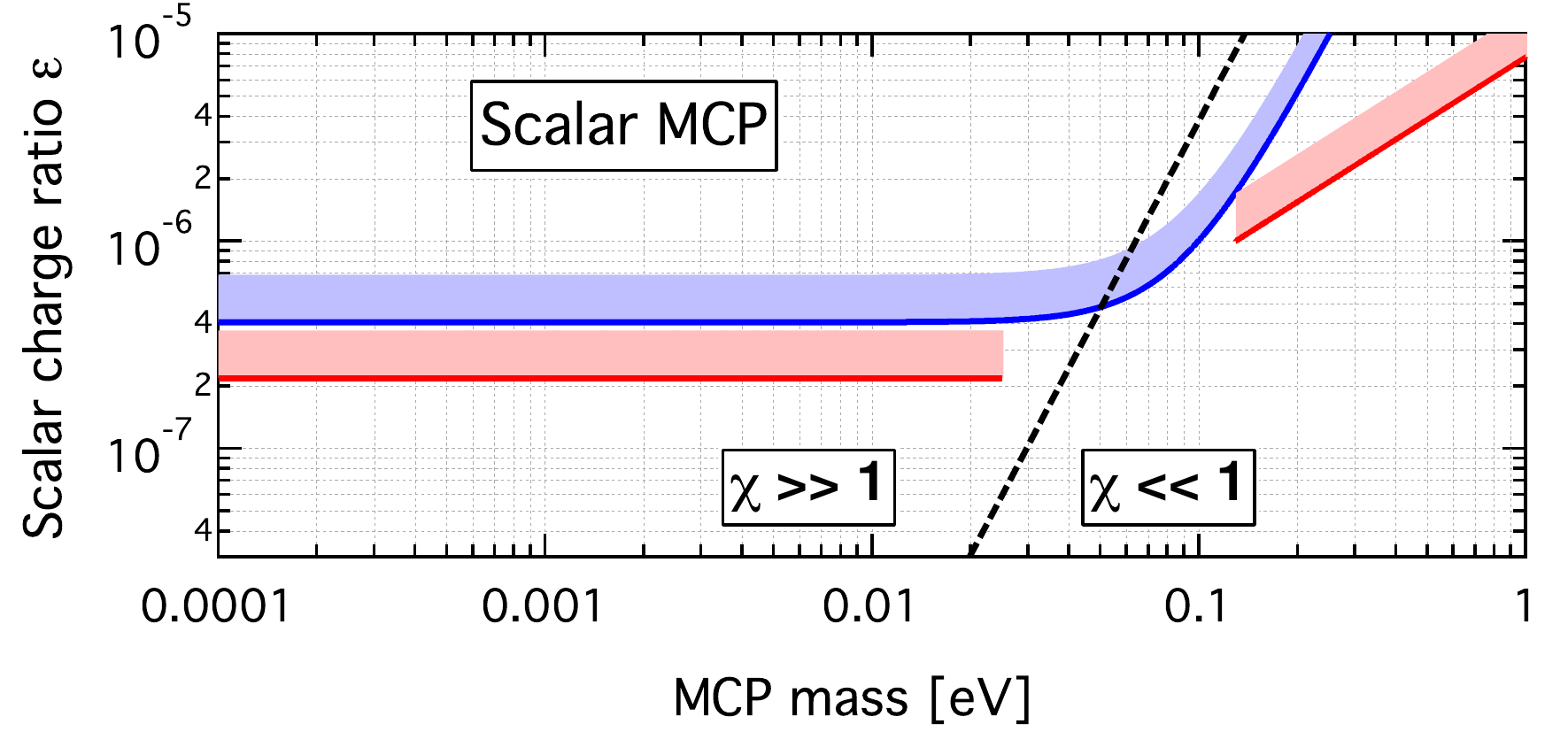}
\caption{Updated 95 \% c.l. exclusion plot for scalar and fermion millicharged particles. In blue is the previous limit taken from \cite{Ahlers} and in red our new limits. The two branches of each of the red ellipticity curves are not connected in the mass range around $\chi\simeq1$ (dotted black line), where the birefringence changes sign.}
\label{mcp}
\end{figure}

In figure \ref{mcp} we show our new limit on $\epsilon$ as a function of particle mass compared to a previous limit obtained from magnetically induced dichroism measurements \cite{Ahlers}. In the case of fermions, this includes neutrinos for which $\epsilon \lesssim 10^{-7}$ for masses below 20 meV. For previous limits see \cite{PDG}.

\section*{Acknowledgements}
We greatly thank Luca Landi for his invaluable technical help during the construction of the apparatus.

\end{document}